\documentclass[a4paper]{article}

\usepackage{INTERSPEECH2020}
\usepackage{comment}
\usepackage{multirow}
\usepackage{tikz}
\usepackage{pgfplots}
\usetikzlibrary{pgfplots.groupplots}
\usetikzlibrary{plotmarks}
\usetikzlibrary{calc}

\title{Massively Multilingual ASR: 50 Languages, 1 Model, 1 Billion Parameters}
\name{Vineel Pratap$^1$, Anuroop Sriram$^1$, Paden Tomasello$^1$, Awni Hannun$^2$, Vitaliy Liptchinsky$^1$, \\ Gabriel Synnaeve$^2$\textsuperscript{\textsection}, Ronan Collobert$^1$\textsuperscript{\textsection}}
\address{
  $^1$Facebook AI Research, Menlo Park\\
  $^2$Facebook AI Research, NYC}
\email{\{vineelkpratap,anuroops,padentomasello,awni,vitaliy888,gab,locronan\}@fb.com}

\def\bx{{\mathbf x}}
\def\by{{\mathbf y}}
\def\bh{{\mathbf h}}
\def\ba{{\mathbf a}}
\def\ll{{\mathcal L}}
\def\gg{{\mathcal G}}
\def\xx{{\mathcal X}}
\def\yy{{\mathcal Y}}

\begin{document}

\maketitle

\begingroup\renewcommand\thefootnote{\textsection}
\footnotetext{Equal Advising}
\endgroup

\begin{abstract}

We study training a single acoustic model for multiple languages with the aim of improving automatic speech recognition (ASR) performance on low-resource languages, and overall simplifying deployment of ASR systems that support diverse languages. We perform an extensive benchmark on 51 languages, with varying amount of training data by language (from 100 hours to 1100 hours). We compare three variants of multilingual training from a single joint model without knowing the input language, to using this information, to multiple heads (one per language ``cluster''). We show that multilingual training of ASR models on several languages can improve recognition performance, in particular, on low resource languages.  We see 20.9\%, 23\% and 28.8\% average WER relative reduction compared to monolingual baselines on joint model, joint model with language input and multi head model respectively. To our knowledge, this is the first work studying multi-lingual ASR at massive scale, with more than 50 languages and more than 16,000 hours of audio across them.


\end{abstract}
\noindent\textbf{Index Terms}: speech recognition, multilingual

\section{Introduction}

The use of multilingual ASR systems~\cite{burget2010multilingual, lin2009study,heigold2013multilingual,bourlard2011current,cho2018multilingual,toshniwal2018multilingual,Kannan_2019,li2019bytes} that simultaneously transcribe multiple languages has recently become popular to increase language coverage, but covering all of the world's $\sim$7000 languages is still far ahead.
The ability to train a single model on many, say more than 50 languages, presents a few advantages. First, in a production setting, training, deploying and maintaining one model per language, especially on long tail of low resource languages, can quickly become cumbersome as the number of languages increases. Having a single model for all languages can simplify the production pipeline significantly. Second, as previously shown in the literature, training multilingual ASR models on a small set of similar languages can improve recognition performance. However, it is not clear if these multi-lingual approaches can scale to a large number of diverse languages, from different language families.


In this work, we (i) train at scale on 51 languages from several language families,
and (ii) show that a joint model with shared vocabulary approach can surpass strong monolingual baselines on low resource languages. Furthermore, (iii) we propose a refined multi-head approach, where each head addresses a set of similar languages and improves on the monolithic joint model approach, leading to competitive results (compared to monolingual baselines) also on higher-resource languages. Finally, (iv) we demonstrate that our multilingual model learns representations general enough that it improves monolingual baseline WER on new languages not seen during the initial training phase.

The rest of the paper is organized as follows. Section 2
presents related work on multilingual ASR. In Section 3,
we review the multilingual models used in our work. In Section 4, we discuss our experimental setup and present our results and analysis in Section 5. Finally, in Section 6 we discuss future work and conclude.

\section{Related Work}

A single model capable of recognizing multiple languages has been a long-term goal in the field of automatic speech recognition. Models capable of learning from multiple languages have been studied in both the HMM-GMM~\cite{burget2010multilingual, lin2009study} and DNN-HMM hybrid systems~\cite{heigold2013multilingual}. In general, multi- and cross-lingual speech processing has been an active area of research for decades~\cite{bourlard2011current}. 

More recently, as end-to-end models have matured in the monolingual setting \cite{Park_2019,synnaeve2019end}, attention has turned to leverage multiple languages to further improve their performance, specifically in the low resource setting. End-to-end models typically require more data to match and surpass the performance of hybrid systems and thus leveraging data from multiple languages is more relevant now than ever. 

Multi-lingual sequence-to-sequence models have been shown to improve performance in the cross-lingual setting~\cite{cho2018multilingual}, where the model is first pre-trained on a group of languages and then fine-tuned to a specific target language. Prior work has leveraged as many as 100 languages simultaneously to learn language agnostic features~\cite{adams-etal-2019-massively}. The work, however, studied a limited dataset which consists of Bible readings in different languages, with only a single reading of Bible for most languages. 

In \cite{toshniwal2018multilingual} and \cite{Kannan_2019}, authors train a sequence-to-sequence and RNN-T~\cite{graves2012sequence} models respectively on 9 Indian languages. Both approaches perform analysis using shared encoder and decoder with language identification as an additional input and train with unified grapheme set from all languages. While the former approach trains a single model, the latter one uses additional adaptive per-language layers. 

In \cite{li2019bytes}, an acoustic model outputs Unicode bytes directly rather than letters or sentence pieces. They show gains in the multi-lingual setting and espouse the efficiency benefits of predicting bytes in avoiding large Softmax layers.

All of the discussed related work either study limited set of languages, typically less than 10, or limited data, such as readings of Bible. To the best of our knowledge, this work is the first one to study multilingual systems at a massive scale, covering 51 languages and more than 16,000 hours of audio.

\begin{figure*}
\centering
\begin{tikzpicture}
\begin{axis}[
    width=17cm, 
    height=4cm,
    axis x line*=bottom,
    axis y line*=left,
    xtick={0,1,2,3,4,5,6,7,8,9,10,11,12,13,14,15,16,17,18,19,20,21,22,23,24,25,26,27,28,29,30,31,32,33,34,35,36,37,38,39,40,41,42,43,44,45,46,47,48,49,50,0,1,2,3,4,5,6,7,8,9,10,11,12,13,14,15,16,17,18,19,20,21,22,23,24,25,26,27,28,29,30,31,32,33,34,35,36,37,38,39,40,41,42,43,44,45,46,47,48,49,50},
    xticklabels from table={data/train_data.dat}{l},
    xticklabel style={align=center,tick label style={rotate=90},font=\scriptsize},
    xmin=-1,
    xmax=51,
    ytick={0,100,500,1000},
    ylabel={Training data (in hrs)},
    y label style={at={(axis description cs:-0.025,.45)},font=\footnotesize},
    yticklabel style={font=\scriptsize},
    legend image code/.code={
        \draw [#1] (0cm,-0.1cm) rectangle (0.4cm,0.2cm); },
    legend style={at={(0.65,0.95)}, anchor=north,legend columns=-1,column sep=0.2cm},
    every axis plot/.append style={
          ybar,
          bar width=2.5pt,
          bar shift=0pt,
          fill
        }
    ]
    \addplot[fill=purple] table[x=x, y=y, restrict y to domain=500:1200] {data/train_data.dat};
    \addplot[fill=teal] table[x=x, y=y, restrict y to domain=300:500] {data/train_data.dat};
    \addplot[fill=violet] table[x=x, y=y, restrict y to domain=0:200] {data/train_data.dat};
    \legend{High Resource, Mid Resource, Low Resource}
\end{axis}
\end{tikzpicture}
\vspace{-8pt}
\caption{Training data distribution across different languages}
\label{img:train_data}
\end{figure*}
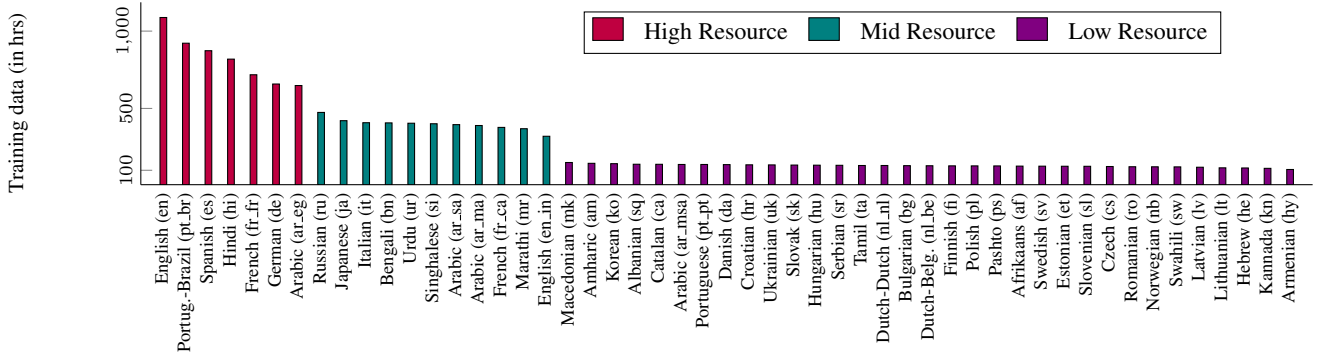

\section{Multilingual models}
\label{sec:mling_models}

\subsection{Seq2Seq Model}

A sequence to sequence (Seq2Seq) model comprises of two neural networks: an encoder and a decoder. The encoder maps the input audio sequence $\bx = (x_1, ..., x_T)$ to an intermediate representation $\bh = (h_1, ..., h_K)$. The decoder maps $\bh$ to the output sequence $\by = (y_1, ..., y_L)$ in an autoregressive manner. Specifically, we use a stacked unidirectional RNN that computes the probability of the sequence $\by$ using:

\begin{equation}
    P(\by | \bh) = \prod_{t} P(y_t | \bh, \by_{<t})
\end{equation}

Thus, the decoder learns a language model conditioned on the hidden representations $\bh$. The encoder and decoder networks are jointly optimized to minimize the cross-entropy loss between the output of the network and the ground truth transcriptions. In our models, the encoders are based on the time-depth separable convolution architecture \cite{Hannun2019SequencetoSequenceSR}.

For multilingual training, we consider $N$ languages $(\ll_1, ..., \ll_N)$ with each language  $\ll_i$ consisting of an independent training set $\{\xx_i, \yy_i\}$ comprising of $n_i$ samples. Each language $\ll_i$ has a set of graphemes $\gg_i$ that may overlap with the graphemes from other languages. We train all of our multilingual models on the combined training set $(\xx, \yy) = \cup_{i=1}^{N} (\{\xx_i, \yy_i\})$. 

\subsection{Shared sub-word tokens} 
\label{sec:subwords}
Working with a large set of languages, each with their own distinct character set and tokenization rules, makes training and maintaining models cumbersome. Adding or removing languages would require modifications to the model architecture and training routine for example. To simplify this process, we create a shared token set across all languages using a Sentence Piece Model (SPM)~\cite{kudo2018subword} . Similar to \cite{conneau2019cross}, the shared sentence pieces are built by sampling the sentences using a multinomial distribution $\{s\}_{i=1..N}$,
\begin{equation}
   s_i = \frac{p_i^\alpha}{\sum_{j=1}^{N}  p_j^\alpha}  \quad   \text{with} \quad  p_i = \frac{n_i}{\sum_{k=1}^{N}  n_k}\,,
\end{equation}
where the parameter $\alpha$ controls the sampling of languages with different frequencies. 

\subsection{Joint model}
Our joint model approach is a single model which is trained while sharing the parameters from the encoder, decoder and token set, across all languages. We optionally (see Section~\ref{sec:results}) feed language information to the model in the form of an embedding, which is also trained jointly with the ASR model. 

\subsubsection{Curriculum training of joint model}
\label{sec:curriculum}
We faced convergence issues with joint model, when training on data from all languages. For these cases, we introduced a curriculum training~\cite{bengio2009curriculum} based approach, which incrementally adds each language after the model has been trained for a fixed number of iterations or the Character Error Rate (CER) goes below 50\% for the previously added language. We found that training converges easily for up to 51 languages using this method. 

\subsection{Multi-headed model}
\label{sec:multihead}
Joint training of multiple tasks can only be beneficial when the individual tasks share common representations. Since the decoder of a sequence-to-sequence model learns a conditional language model, sharing decoder parameters between languages that do not have any graphemes in common is unlikely to improve the recognition performance of any of the languages. Therefore, we divide the languages into $M$ distinct groups and use a different decoder for each language group. Thus, our multi-headed models employ a single encoder whose parameters are shared across all languages, and a set of $M$ decoders, one per language group. We select 10K subword units as the token set for each language group as described in section \ref{sec:subwords}. In the forward pass, the appropriate decoder is selected based on the language.

Ideally, we would like to group the languages by their written scripts. However, this leads to a skewed distribution of group sizes, with a few language groups containing many languages and others containing only a single language. In such a setting, it becomes critical to tune the decoder hyperparameters (like the number of RNN layers) for each language group, adjusting the head capacity according to the amount of training data available for that group. To avoid this additional complication, we manually combined some of the smaller language groups together until we end up with six language groups. The language groups we used in our experiments are shown in Table~\ref{tab:langgroups}. We do not use curriculum training (Section~\ref{sec:curriculum} for multi-headed models because they are able to converge even when trained with all 51 languages together.

\begin{table}[t]
    \centering
    \begin{small}
    \begin{tabular}{r|l}
        \toprule
        Group name & Languages \\
        \midrule
        Latin & af, ca, da, de, en, en\_in, es, et, \\
              & fi, fr\_ca, fr\_fr, hu, it, lt, nl\_be, nl\_nl, pt\_br, \\
              & pt\_pt, ro, sq, sv, sw \\
        Balto-Slavic & cs, hr, lv, nb, pl, sk, sl \\
        Indic & bn, hi, kn, mr, si, ta \\
        Perso-Arabic & am, ar\_eg, ar\_ma, ar\_msa, ar\_sa, he, ps, ur \\
        Cyrillic & bg, mk, ru, sr, uk \\
        Misc & hy, ja, ko \\
    \bottomrule
    \end{tabular}
    \caption{Language groups used for the multi-headed models}
    \label{tab:langgroups}
    \end{small}
\end{table}

\section{Experimental details} 

\subsection{Data}
The training set used for our experiments consists of videos publicly shared by users that are anonymized and spans a total of 51 languages. Figure \ref{img:train_data} shows the amount of training data present in all the languages. We categorize the languages into three categories -- high resource languages consisting of \textgreater 600 hours of train data, mid resource language with 300-500 hours of training data and low resource languages with 100-150 hrs of training data. We use about 20 hours of test set for each language. All hyper parameter tuning is done on a held-out development set of about 13 hours of high and mid resource languages, and about 7 hours for low resource languages.

\subsection{Data preprocessing}
Since our dataset is transcribed with predefined guidelines, we were able to avoid many nuances which can arise when mining the text from online sources. For each language, we normalize the text by performing NFKC normalization and removing all punctuations. We then prepare a list of valid unicode characters based on the language's orthography and filter words which contain characters outside this range. We use this data for generating the token set and lexicon as well as model training.

\subsection{Training setup}
All our experiments are run using wav2letter++~\cite{pratap2018} framework. We use 80-dimensional log mel-scale filter banks as input features, with STFTs computed on 30ms Hamming windows strided by 10ms. All our acoustic models are based on the system proposed in \cite{Hannun2019SequencetoSequenceSR}. We use SpecAugment \cite{Park_2019} for all our experiments with LibriSpeech Double setting. We also use Blockwise Momentum Update Filtering (BMUF) \cite{7472805} for all the experiments to help with scaling the training workflows. As local criterion in BMUF, we use Stocastic Gradient Descent (SGD) with momentum. 

\subsection{Monolingual baseline models}
All baseline models use an encoder with three 10-channel, four 14-
channel and eight 18-channel Time Depth Separable Convolution (TDS) \cite{Hannun2019SequencetoSequenceSR}  blocks. We use three 1D convolutions to sub-sample over time, one as the first layer and one in between each group of TDS blocks. Each 1D convolution module has a stride of 2 which accounts for a total sub-sampling factor of 8. We use kernel size of 21 for all the convolution layers. These layers are followed by a final Linear layers which produces 1024-dimensional encoder output. The decoder which is also based on \cite{Hannun2019SequencetoSequenceSR} consists of two-layer GRU with 512 hidden units with two rounds of inner-product key-value attention.  Overall, the combined encoder and decoder model has about 150 Million parameters. 

We have tuned dropout and hyper parameters in BMUF extensively for all the models. For the high and mid resource languages, we use 5000 and 2000 sub-word tokens respectively generated from SentencePiece toolkit~\cite{kudo2018subword}. For low resource languages, we use graphemes as the modelling units as it gave better performance over sub-word units. For all languages, the test WER is taken for the epoch which produces the best validation WER. 


\subsection{Training data sampling for multilingual models}
\label{sub_sec:data_sampling}
Because of the imbalance of data across languages,  it can be difficult for models to perform well on low resource languages. Similar to \cite{Kannan_2019}, we sample data from a language $\ll_i$ during training from a multinomial distribution ${s}_{i=1..N}$ as given below
\begin{equation}
s_i = \frac{n^{max} + \beta * (n_i - n^{max})}{\sum_{j=1}^{N}  n^{max} + \beta * (n_j - n^{max} )}    
\end{equation}
 where $n^{max}$ is the maximum number of training samples across any language and  $\beta$ is a tunable parameter that allows us to adjust the sampling of languages from their natural frequency, when $\beta$ = 1, to a uniform distribution across languages when $\beta$ = 0.  

\section{Results and analysis}
\label{sec:results}

In this section, we present our study on three multilingual models - joint model, joint model with language input and multi-head model described  in Section \ref{sec:mling_models}.

\subsection{Study of tunable parameters $\alpha$, $\beta$}

\begin{figure}[h!]
\centering     
\includegraphics[width=\linewidth]{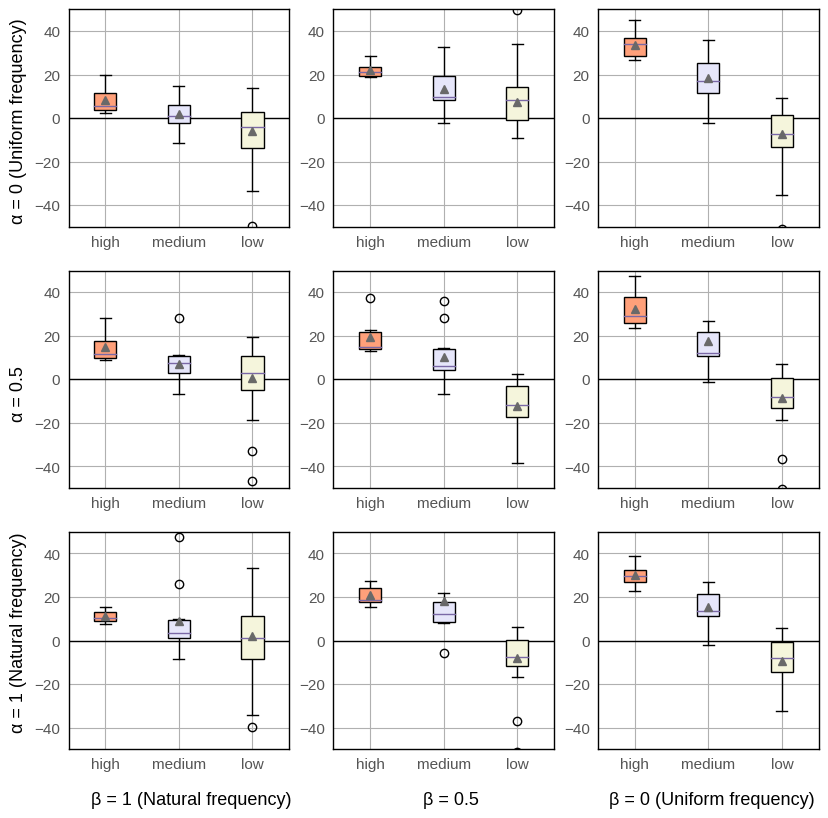}
\caption{Box plots of relative WER change from monolingual baseline (lower is better) on a 500 million joint model using 10K shared sentence pieces with varying tunable parameters $\alpha$ and $\beta$. $ \alpha = 0$ and $\beta = 0 $  represents sampling from all languages uniformly for both vocabulary creation and multilingual model training. $\alpha = 1 $ and $ \beta = 1$ represents sampling from all languages at their natural frequency. Languages are sectioned by their resource category.}
\label{img:sampling_ablation}
\end{figure}

As mentioned in section \ref{sec:subwords} and \ref{sub_sec:data_sampling}, we use tunable parameters $\alpha$, $\beta$ for controlling the sampling of languages during token generation and training examples during multilingual model training respectively. We compare the WER performance of a 500 million parameter joint model with varying sampling fractions and the results are presented in Figure~\ref{img:sampling_ablation}.


In general, we see that going from natural frequency ($ \alpha = 1 $, $ \beta = 1$) to uniform frequency  ($ \alpha = 0 $, $ \beta = 0$) seems to improve performance of low resource languages while degrading performance on high resource languages. Interestingly, it appears the using a $ \alpha = 0.5 $ and $ \beta = 0.5$ performs best on low resource languages and has less performance degradation on high, mid resource languages compared to sampling at uniform frequency ($ \alpha = 0 $ and $ \beta = 0$). For low resource language, one might assume that sampling a language more frequently will always result in better performance. We believe that sampling at the natural frequency has too much data imbalance to learn an effective shared representation, while sampling at the uniform distribution overfits to the low resource languages. We use $ \alpha = 0.5 $ and $ \beta = 0.5$ for all of our multilingual experiments.

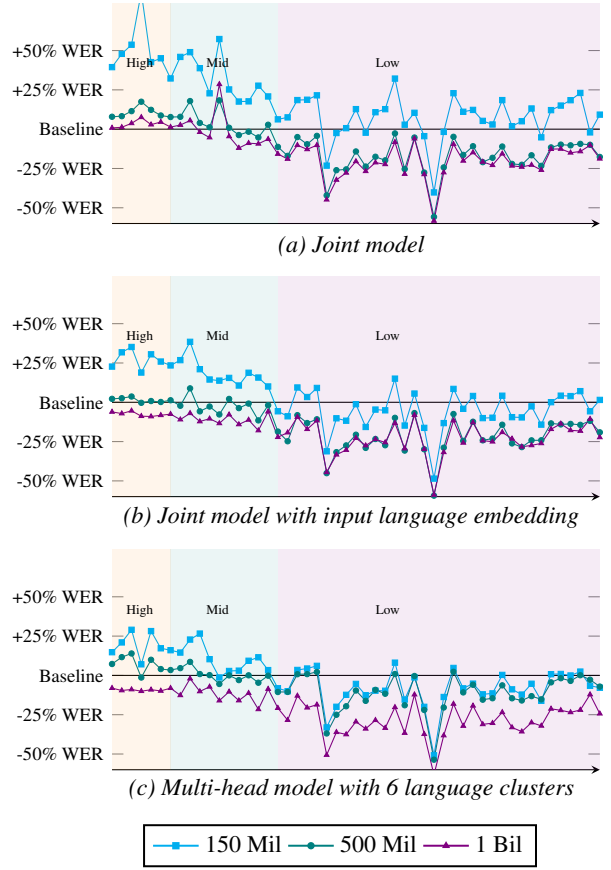
\begin{figure}[!ht]
\begin{tikzpicture}
\tikzset{mark options={mark size=2}}
\begin{groupplot}[
    group style={
        group name=my plots,
        group size=1 by 3,
        xticklabels at=edge bottom,
        vertical sep=20pt,
    },
    width=8cm,
    height=4.5cm,
    axis x line=bottom,
    y axis line style={draw opacity=0},
    yticklabel style={font=\scriptsize, draw=none},
    y label style={at={(axis description cs:0.043,.45)},font=\footnotesize},
    xlabel={High$\quad\quad\quad\quad$Mid$\quad\quad\quad\quad\quad\quad\quad\quad\quad\quad\quad$Low}, 
    x label style={at={(axis description cs:0.31,0.97)},font=\tiny},
    ymin=-60,
    ymax=80,
    xtick=\empty,
    ytick={ 50, 25, 0, -25, -50},
    yticklabels={ +50\% WER, +25\% WER, Baseline, -25\% WER, -50\% WER},
]
\nextgroupplot[legend style={at={($(0,0)+(3cm,-8.25cm)$)},legend columns=3,fill=none,draw=black,anchor=center,align=center,line width=0.75pt}]
\addplot[forget plot, mark size=0] table [x=x, y=b] {data/final_wer.dat};
\addplot[mark=square*,color=cyan ,mark options={scale=0.5}] table [x=x, y=c] {data/final_wer.dat};
\addplot[mark=*,color=teal,mark options={scale=0.5}] table [x=x, y=e] {data/final_wer.dat};
\addplot[mark=triangle*,color=violet ,mark options={scale=0.5}] table [x=x, y=f] {data/final_wer.dat};
\addplot[fill=orange, fill opacity = 0.08, draw = none,area legend] coordinates {(0,-200) (0,200) (6,200)(6,-200)};
\addlegendentry{150 Mil};
\addplot[fill=teal, fill opacity = 0.08, draw = none,area legend] coordinates {(6,-200) (6,200) (17,200)(17,-200)};
\addlegendentry{500 Mil};
\addplot[fill=violet, fill opacity = 0.08, draw = none,area legend] coordinates {(17,-200) (17,200) (50,200)(50,-200)};
\addlegendentry{1 Bil};
\nextgroupplot
\addplot[mark size=0] table [x=x, y=b] {data/final_wer.dat};
\addplot[mark=square*,color=cyan ,mark options={scale=0.5}] table [x=x, y=g] {data/final_wer.dat};
\addplot[mark=*,color=teal,mark options={scale=0.5}] table [x=x, y=i] {data/final_wer.dat};
\addplot[mark=triangle*,color=violet ,mark options={scale=0.5}] table [x=x, y=j] {data/final_wer.dat};
\addplot[fill=orange, fill opacity = 0.08, draw = none] coordinates {(0,-200) (0,200) (6,200)(6,-200)};
\addplot[fill=teal, fill opacity = 0.08, draw = none] coordinates {(6,-200) (6,200) (17,200)(17,-200)};
\addplot[fill=violet, fill opacity = 0.08, draw = none] coordinates {(17,-200) (17,200) (50,200)(50,-200)};
\nextgroupplot
\addplot[mark size=0] table [x=x, y=b] {data/final_wer.dat};
\addplot[mark=square*,color=cyan ,mark options={scale=0.5}] table [x=x, y=k] {data/final_wer.dat};
\addplot[mark=*,color=teal,mark options={scale=0.5}] table [x=x, y=m] {data/final_wer.dat};
\addplot[mark=triangle*,color=violet ,mark options={scale=0.5}] table [x=x, y=n] {data/final_wer.dat};
\addplot[fill=orange, fill opacity = 0.08, draw = none,area legend] coordinates {(0,-200) (0,200) (6,200)(6,-200)};
\addplot[fill=teal, fill opacity = 0.08, draw = none,area legend] coordinates {(6,-200) (6,200) (17,200)(17,-200)};
\addplot[fill=violet, fill opacity = 0.08, draw = none,area legend] coordinates {(17,-200) (17,200) (50,200)(50,-200)};
\end{groupplot}
\node (title) at (3.15, -0.3) {\textit{(a) Joint model}};
\node (title) at (3.15, -3.9) {\textit{(b) Joint model with input language embedding}};
\node (title) at (3.15, -7.5) {\textit{(c) Multi-head model with 6 language clusters}};
\end{tikzpicture}
 \caption{Relative WER change (lower is better) for different multilingual models as we increase model size. The amount of training data is gradually reducing as we move along x-axis for each plot. For 'ja' and 'ko', we use Character Error Rate (CER) instead of Word Error Rate (WER)}
 \label{fig:all_wers}
\end{figure}
    

\subsection{Joint model}

We use a shared token set of 10K sentence pieces for all joint model experiments. We have also tried joint models with shared graphemes and shared sentence pieces of size 25K and 50K and empirically found that 10K sentence pieces give the best performance. For the joint model with input language embedding, we use a 10-dimensional language embedding and concatenate it with the 80-dimensional log-mel input features and feed it to the encoder. 

 As mentioned in Section~\ref{sec:curriculum}, we use curriculum training for training the models. SpecAugment\cite{Park_2019} is applied once all the languages added in the curriculum training. Figure \ref{fig:all_wers}(a)  shows the results of joint model and Figure \ref{fig:all_wers}(b) shows the results of joint model with input language embedding for different model sizes. We can see that increasing the model size helps with improving WER in both settings.  
 
 For 1 billion parameter joint model, we see an average WER degradation of 3.15\% on high resource and an average WER improvement of 2.5\% and 20.87\% on mid and low resource languages. Further, we observe that using language embedding at the input layer performs better that without using it for a given model size. For 1 billion parameter joint models with language embedding, we observe an improvement in WER on all the languages. The average WER improvements on high, mid and low resource languages are 7.48\%, 12.11\% and 23.03\% respectively. 
 
\subsection{Multi-head model} 

Figure \ref{fig:all_wers}(c) shows the relative change in WER obtained with multi-headed models of different sizes compared to the baseline for each language. The largest multi-headed model with 1 billion parameters can significantly improve performance on all languages. This model improves WER by 9.1\% on average for high-resource languages, by 12.44\% for mid-resource languages, and by 28.76\% for low-resource languages. The largest multi-headed model also outperforms the joint models even when the joint models are fed a language embedding. In addition, the multi-headed models are simpler to train as they do not need curriculum training.

\subsection{Multilingual transfer learning on unseen languages}
Training multilingual models on a large, diverse set of languages enables the acoustic models to learn language-agnostic representations general enough to perform well on completely new languages. To demonstrate this, we fine-tune the joint model with 1 Billion parameters on 3 unseen low-resource languages (100-150 hrs of training data). Since, the graphemes in new languages, which are being fine-tuned, are not present in the decoder of trained joint model, we re-initialize the decoder for the grapheme set of new language. We allow both encoder and decoder to be trained during fine-tuning. 

From table \ref{tab:fine_tune}, we can see that fine-tuning on multilingual joint model improves the WER over monolingual baselines. The fine-tuning approach can thus help with adapting a new language easily while also improving the WER from monolingual baseline.  

\begin{table}[h!]
\centering
 \begin{small}
  \begin{tabular}{lcc}
    \toprule
    \textbf{Language}     & \textbf{Monolingual }  & \textbf{Fine-tuning on}       \\
    \textbf{}     & \textbf{ Baseline}  & \textbf{Joint Model}   \\
    \midrule
	Chinese (zh\_tw) & 50.82 & 39.29 (-22.7\%) \\
    Persian (fa) & 33.59 & 31.29 (-6.8\%) \\
	Telugu (te) & 50.05 & 47.63  (-4.8\%)\\

\midrule
  \end{tabular}
    \caption{Fine-tuning results on joint model. For ’zh\_tw’, we report Character Error Rate (CER) instead of Word Error Rate (WER}
      \label{tab:fine_tune}
   \end{small}
\end{table}



\begin{figure}
\centering     
\includegraphics[height=6cm]{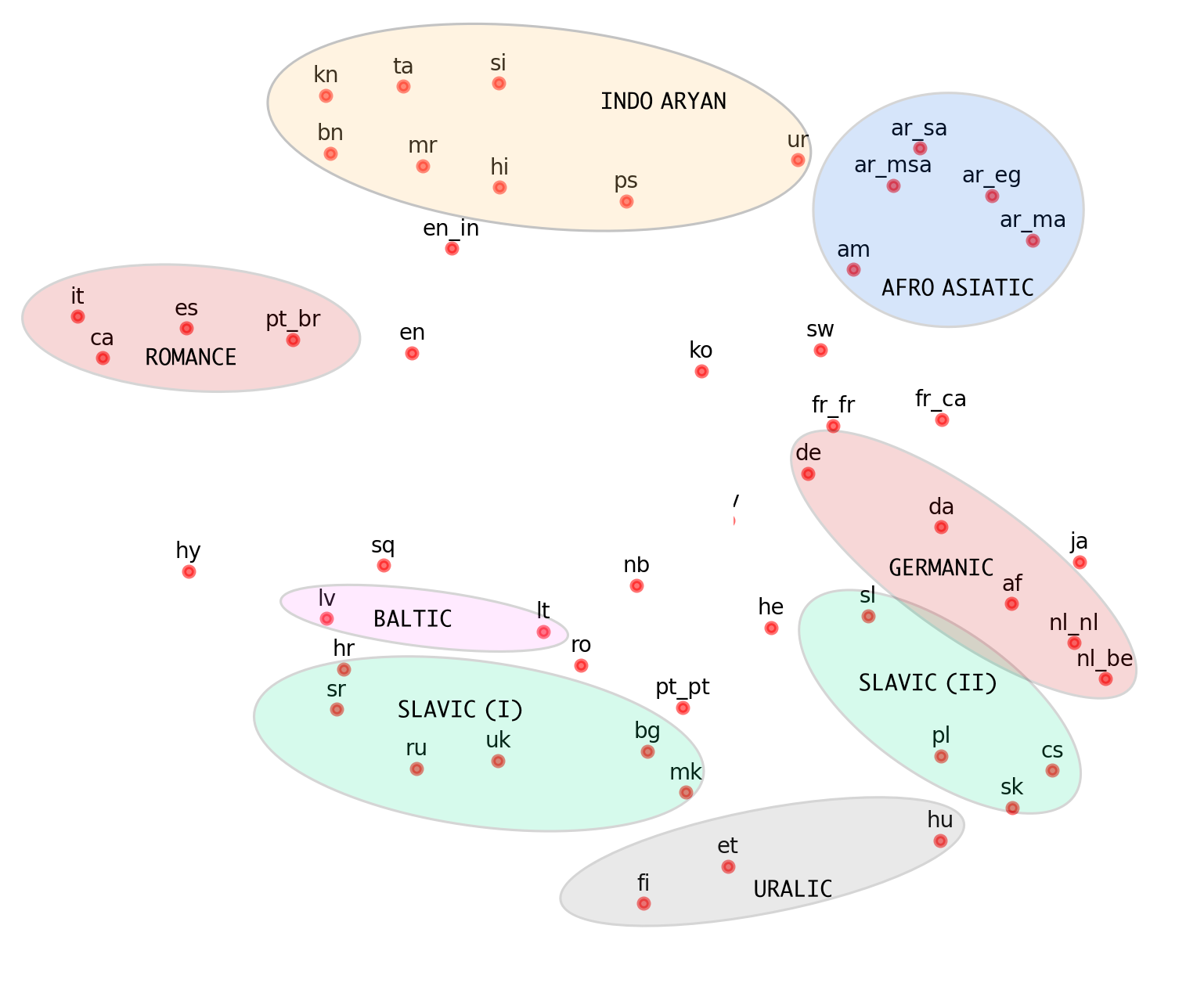}
\caption{t-SNE clustering of trained language embeddings. Colored clusters are based on language families.}
\label{img:tsne}
\end{figure}

\subsection{Language embedding analysis}
We use t-SNE \cite{maaten2008visualizing} method to visualize the learned embeddings of the joint model trained with input language embedding. From Figure \ref{img:tsne}, we can notice the learned language embeddings form noticeable clusters for language families. Similar to \cite{tan2019multilingual}, these learned clusters can be used for training multi-head experiments instead of choosing the clusters manually, but we will leave it for future work. 

\section{Conclusion}

We demonstrated that it is possible to train a massive single ASR architecture for 51 various languages, which we found in practice considerably less time-consuming to tune than 51 different monolingual baselines. 

\section{Acknowledgements}
We would like to thank Steven Garan for help in data preparation and text normalization.

\bibliographystyle{IEEEtran}

\bibliography{mybib}

\end{document}